\documentclass[a4paper,american]{scrartcl}

\usepackage{tikz-cd}
\usepackage{verbatim}
\usepackage{enumerate}
\usepackage{lmodern}
\usepackage[utf8]{inputenc}
\usepackage[T1]{fontenc}
\usepackage{color}
\usepackage{todonotes}
\usepackage{babel}
\usepackage{amsfonts, amsmath, amsthm, amssymb}
\usepackage{setspace}
\usepackage[authoryear]{natbib}
\usepackage[font=small]{quoting}
\usepackage{latexsym}
\usepackage[bookmarks=false,pdfborder={0 0 0},colorlinks=false]{hyperref}
\usepackage{dsfont}
\usepackage{xcolor}
\makeatletter

\usepackage[arrow, matrix, curve]{xy}

\newcommand{\IP}{\mathbb P}

\newcommand{\dd}{\mathrm{d}}

\newcommand{\supp}{\mathrm{supp}\,}

\theoremstyle{plain}

\theoremstyle{definition}

\makeatother

\begin{document}

\title{Observables and unobservables in quantum mechanics: How the no-hidden-variables theorems support the Bohmian particle ontology}

\author{Dustin Lazarovici%
\thanks{Université de Lausanne, Section de philosophie,
1015 Lausanne, Switzerland. E-mail:
\protect\href{mailto:dustin.lazarovici@unil.ch}{dustin.lazarovici@unil.ch}}%
, Andrea Oldofredi%
\thanks{Université de Lausanne, Section de philosophie,
1015 Lausanne, Switzerland. E-mail: \protect\href{mailto:Andrea.Oldofredi@unil.ch}{Andrea.Oldofredi@unil.ch}%
}, Michael Esfeld%
\thanks{Université de Lausanne, Section de philosophie,
1015 Lausanne, Switzerland. E-mail: \protect\href{mailto:Michael-Andreas.Esfeld@unil.ch}{Michael-Andreas.Esfeld@unil.ch}}
}

\maketitle
\begin{abstract}
forthcoming in \emph{Entropy}, special issue Emergent Quantum Mechanics -- David Bohm Centennial Perspectives
\vspace{2mm} 

The paper argues that far from challenging -- or even refuting -- Bohm's quantum theory, the no-hidden-variables theorems in fact support the Bohmian ontology for quantum mechanics. The reason is that (i) all measurements come down to position measurements and (ii) Bohm's theory provides a clear and coherent explanation of the measurement outcome statistics based on an ontology of particle positions, a law for their evolution and a probability measure linked with that law. What the no-hidden-variables theorems teach us is that (i) one cannot infer the properties that the physical systems possess from observables and that (ii) measurements, being an interaction like other interactions, change the state of the measured system.    
\medskip{}
\vspace{2mm} 

\end{abstract}

\tableofcontents{}

\section{Introduction}

The famous no-hidden-variables theorems have played a crucial, though often questionable role in the history of quantum mechanics. For decades they have been employed to defend the quantum orthodoxy and to argue, \emph{nay prove}, that any attempt to go beyond the statistical formalism of standard quantum mechanics in providing a ``complete'' description of the microcosm is bound to fail. Even after David \cite{Bohm:1952aa} got ``the impossible done'' (as \citet[][p. 160]{Bell:2004aa} later put it) and showed how the statistical predictions of quantum mechanics can be derived from an ontology of point particles and a deterministic law of motion, many scientists and philosophers refused to pay attention to this theory on the basis that the no-hidden-variables theorems had established that it couldn't be correct (one striking example of such a misunderstanding is \citet[][pp. 53-55]{Wigner:1983}).

Of course, Bohm's theory is not a counterexample to these theorems qua mathematical theorems. It is rather the most striking demonstration of the fact that these mathematical results do not support the ideological conclusions in defense of which they have been generally cited. That notwithstanding, it would be premature to dismiss the ``no-go theorems'' as physically and philosophically irrelevant. They capture something not only about the nature of measurements and the statistical predictions of quantum mechanics that strikes us as remarkable and contrary to classical intuitions, but also about the nature of physical objects. The aim of this paper is to work out what exactly these theorems show and how they support in fact Bohm's quantum theory, instead of being an argument against it.

In the next section, we briefly recall the quantum orthodoxy and Bohm's quantum theory. Section 3 outlines three of the most important theorems useful for our discussion. Section 4 rebuts the conclusions that are commonly drawn from them. Section 5 provides an account of the Bohmian theory of measurements. Section 6 shows how it supports an ontology of point particles that are characterized by their positions only. Section 7 draws a general conclusion.

\section{Quantum orthodoxy and Bohmian mechanics}\label{sec:why}

In the words of David \citet[][p. 803]{Mermin:1993aa}, the scope of the no-hidden-variables theorems is to defend ``a fundamental quantum doctrine'', namely that

\begin{itemize}
\item[(Q)] A measurement does not, in general, reveal a preexisting value of the measured property.
\end{itemize}

\noindent However, accepting this doctrine leads to at least two urgent questions: 

\begin{enumerate}
	\item \emph{How do the quantum observables acquire definite values upon measurement?}\\
	It is now generally acknowledged that measurements are not a new type of interaction -- let alone a primitive metaphysical concept -- that requires a special treatment, but come under the common types of physical interactions (electromagnetism, gravitation, etc.). Hence, our physical theories should be able, at least in principle, to describe them. This, in turn, entails that the notion of measurement must not be part of the axioms of a physical theory. Thus, if quantum theory implies that the observable values are not merely revealed but produced by the measurement process -- that is, by the interaction between the measurement device and the measured system --, the theory should tell us how they are produced. 
		
	\item \emph{What characterizes a physical system prior to -- or better: independent of -- measurement?}\\
	After all, there must be some sort of ontological underpinning to the measurement process and the empirical data that it yields. That is, there must be \emph{something} in the world on which the measurement is actually performed -- something with which the observer or measurement device interacts --, and there must be something definite about the physical state of the observer or measurement device that does not, in turn, require a measurement of the measurement (and so on, \emph{ad infinitum}). 
\end{enumerate}

\noindent According to \citet[][p. 803]{Mermin:1993aa}, the orthodox response to question 1 is that ``Precisely how the particular result of an individual measurement is brought into being -- Heisenberg's `transition from the possible to the actual' -- is inherently unknowable''. The response to 2 seems to be some sort of radical idealism, expressed in his now famous assertion (and belated response to Einstein) according to which the moon is demonstrably not there when nobody looks (\cite{Mermin:1985aa}). Bohm's theory entirely rejects this way of talking. Its presentation as a ``hidden variables theory'' suggests that it denies the doctrine Q. However, most contemporary Bohmians actually endorse this doctrine, and quite emphatically so. Let us briefly recall why this is the case.

For present purposes, we use the formulation of Bohm's theory that is today known as Bohmian mechanics (see \cite{Durr:2013aa}; for a discussion of the different contemporary formulations of Bohm's theory, see \cite{Belousek:2003aa}; \cite{Bohm:1993aa} is the latest elaborate treatment by Bohm himself). Bohmian mechanics can be defined in terms of the following four principles:

\begin{enumerate}
\item \emph{Particle configuration}: There always is a configuration of $N$ permanent point particles in the universe that are characterized only by their positions $X_1,\dots,X_N$ in three-dimensional, physical space at any time $t$.
\item  \emph{Guiding equation}: A wave function $\Psi$ is attributed to the particle configuration, being the central dynamical parameter for its evolution. On the fundamental level, $\Psi$ is the \emph{universal} wave function attributed to all the particles in the universe together. The wave function has the task to determine a velocity field along which the particles move, given their positions. It accomplishes this task by figuring in the law of motion of the particles, which is known as the guiding equation:
\begin{equation}\label{guide_eq}
\frac{\mathrm{d}X_k}{\mathrm{d}t}=\frac{\hbar}{m_k}\mathrm{Im}\frac{\nabla_k \Psi}{\Psi} (X_1,\dots,X_N).
\end{equation}
This equation yields the evolution of the $k$th particle at a time $t$ as depending on, via the wave function, the position of all the other particles at that time.
\item \emph{Schrödinger equation}: The wave function always evolves according to the Schrödinger equation:
\begin{equation}
\label{Schr\"odinger_eq}
i\hbar\frac{\partial{\Psi}}{\partial{t}}=-\sum^{N}_{k=1}\frac{\hbar^2}{2m_k}\Delta_k\Psi +V\Psi
\end{equation}
\item \emph{Typicality measure}: On the basis of the universal wave function $\Psi$, a unique\footnote{See \cite{Goldstein:2007aa} for a proof and precise statement.} stationary (more precisely: equivariant) typicality measure can be defined in terms of the $|\Psi|^2$--density. Given that typicality measure, it can then be shown that for nearly all initial conditions, the distribution of particle configurations in an ensemble of sub-systems of the universe that admit of a wave function $\psi$ of their own (known as effective wave function) is a $|\psi|^2$--distribution. A universe in which this distribution of the particles in sub-configurations obtains is considered to be in quantum equilibrium.
\end{enumerate}

\noindent Assuming that the actual universe is a typical Bohmian universe in that it is in quantum equilibrium, one can hence deduce Born's rule for the calculation of measurement outcome statistics on sub-systems of the universe in Bohmian mechanics (instead of simply stipulating that rule). In a nutshell, the axiom of $|\Psi|^2$ providing a typicality measure with $\Psi$ being the universal wave function justifies applying the $|\psi|^2$--rule for the calculation of the probabilities of measurement outcomes on particular sub-systems within the universe, with $\psi$ being the effective wave function of the particular systems in question (see \citet[][ch. 2]{Durr:2013aa}; cf. section \ref{sec:spin} for the notion of effective wave functions). Thus, the quantum probabilities have in Bohmian mechanics exactly the same status as the probabilities in classical statistical mechanics: they are derived from a deterministic law of motion via an appropriate probability measure that is linked with the law. Moreover, if a sub-system admits for an autonomous description in terms of an effective wave $\psi$, its complete physical and dynamical state at any time $t$ is given by the pair $(X_t, \psi_t)$, where $X_t=\left(X_1(t), ...,X_M(t)\right)$ describes the actual spatial configuration of the system. 

Consequently, measurements of observables such as energy, angular momentum, spin, etc. do not reveal predetermined properties of the particles, because Bohmian mechanics does not admit them as intrinsic properties of the particles to begin with. Similarly, a simple analysis of the theory shows that a measurement of the momentum observable does not, except under special circumstances, measure the instantaneous velocity of a particle. It is a crucial feature of the theory that the only property of the particles is their position in space. The particles have a velocity, of course, but velocity is nothing else than the change of position in time. The Bohmian velocity is not an observable (see \citet[][ch. 3.7.2]{Durr:2013aa} for a simple proof, but also \cite{wiseman2007} for the possibility of weak measurements; see \citet[][ch. 7]{Durr:2013aa} for a good discussion of both results). Velocity is not -- in contrast to the Newtonian case -- a dynamical degree of freedom that can be specified independently of the position, because the guiding law \eqref{guide_eq} is a first order differential equation, requiring only positions as initial data.

The first and foremost role of the wave function is a dynamical one, namely to yield the motion of the particles as output, given their positions as input. This explains the name \emph{pilot-wave} theory historically given to Bohm's theory, as if the particles were literally guided or piloted by a wave in physical space. This way of speaking, however, cannot be taken literally, since the wave function is defined on configuration space; it is not a wave propagating in physical space (for the debate about the status of the wave function in Bohmian mechanics, see \cite{Esfeld:2014ab}). Even in the special case when the wave function of a subsystem happens to be an eigenstate $\psi_\alpha$ of a certain observable $\hat A$ with eigenvalue $\alpha$ -- for instance after an ideal measurement -- and it would be safe to say that ``the particle possesses a definite value of $A$'', this way of speaking is unwarranted. It should be replaced by the statement that a (repeated) experiment, whose statistics is encoded in the operator $\hat A$, would yield the outcome $\alpha$ with certainty; or simply by the statement that the effective wave function, guiding the motion of the system, is $\psi_\alpha$. In sum, the validity of doctrine Q is not denied, but substantiated by Bohmian mechanics on the basis of this theory recognizing only position as a property of the physical system. 

\section{No-hidden-variables theorems}

The basic question that the no-hidden-variables theorems set out to address is whether the probabilistic nature of the quantum formalism allows for an ignorance interpretation in the sense that the measured values of quantum observables are in fact predetermined by additional parameters, whose actual values, in individual runs of an experiment, are unknown to us, but whose statistical distributions over a series of measurements reproduce the observed outcome statistics. In more formal terms, the question is whether for any relevant family of quantum observables $\hat{A}, \hat{B}, \hat{C}, \ldots$  there exists a corresponding family of random variables $Z_A, Z_B, Z_C, \ldots$ on a common probability space $\Omega$ such that the values of these random variables correspond to the possible measurement outcomes -- that is, the eigenvalues of the observable operators. Any $\omega \in \Omega$ would then be a value of the  hypothetical hidden variable(s), determining the measurement values $Z_A(\omega), Z_B(\omega), Z_C(\omega), \ldots$, and the quantum predictions, for some quantum state $\psi$, would be reproduced by a probability distribution $\mu_\psi$ over this hidden variable, such as
$\langle \psi \lvert \hat{A} \rvert \psi \rangle = \int_{\Omega} Z_A(\omega) \mathrm{d}\mu_\psi(\omega) $, etc. 

A no-hidden-variables theorem is thus, in general, a result of the following form (cf.  \citet[][ch. 3]{Durr:2013aa}): 

\begin{quote} There is no ``good'' map $\hat{A} \mapsto Z_A$ from the set of self-adjoint operators on a Hilbert space $\mathcal{H}$ to random variables on a common probability space $\Omega$ such that the possible values of $Z_A$ correspond to the eigenvalues of  $\hat{A}$ (that is, the possible measurement values). 
\end{quote}
The term ``good map'' is not quite precise, but deliberately so, for it is essentially on this point -- the requirements on the assignment $\hat{A} \mapsto Z_A$ -- that the various no-hidden-variables theorems differ. 

\subsection{von Neumann} The first no-hidden-variables theorem was proven by von Neumann in his seminal 1932 book \emph{Mathematische Grundlagen der Quantenmechanik} (\cite{Neumann:1932aa}, English translation \cite{Neumann:1955aa}). In this theorem, a ``good'' map from observables to random variables was supposed to be linear, that is, in particular:
\begin{equation}\label{vonNeumannassumption} \hat{A} + \hat{B} \mapsto Z_{A+B}=Z_A+Z_B. \end{equation}
It is easy to see that such a map cannot exist, since for non-commuting operators, the eigenvalues of their sum are in general not sums of their eigenvalues. Von Neumann's linearity assumption was arguably motivated by the additivity of quantum mechanical expectations values ($\langle \psi |\hat{A} + \hat{B}|\psi \rangle = \langle \psi |\hat{A} |\psi \rangle + \langle \psi | \hat{B}|\psi \rangle$ holds for all observables $\hat{A},\hat{B}$ and any state $\psi$), but is nowadays considered as rather naive (\citet[][pp. 805-806]{Mermin:1993aa} calls it ``silly''). As \citet[][p. 806]{Mermin:1993aa} points out, requiring \eqref{vonNeumannassumption} ``is to ensure that a relation holds in the mean by imposing it case by case -- a sufficient, but hardly a necessary condition''. In addition,  the physical significance of this assumption -- in particular for non-commuting observables that cannot even simultaneously measured -- is rather obscure. If, let's say, $\hat X$ is the position and $\hat P$ the momentum observable, what is a ``measurement of $\hat X + \hat P$'' even supposed to mean? For decades, von Neumann's impossibility proof was a key element in the defense of the quantum orthodoxy, but it started to fall apart rather quickly, once people began to study it more systematically. 
 
\subsection{Kochen-Specker}\label{sec:KS}
The theorem of \cite{Kochen:1967aa} was a considerable improvement because it makes a requirement for the ``goodness'' of the assignment $\hat{A} \mapsto Z_A$ that seems \emph{a priori} much more plausible: 

\begin{quote}(NC) \; Whenever the quantum mechanical joint distribution of a set of self-adjoint operators $(A_1, \ldots ,A_m)$ exists, that is, when they form a commuting family, the joint distribution of the corresponding set of random variables, that is, of $(Z_{A_1},\ldots  ,Z_{A_m} )$, must agree with the quantum mechanical joint distribution.\end{quote} 

\noindent This assumption actually implies that all algebraic identities which hold between the observable operators must also hold between the random variables,\footnote{E.g. if $\hat A\cdot\hat B =\hat B\cdot\hat A = \hat C$, it means that the joint distribution is zero on the value set $\lbrace(c\neq ab)\mid a,b,c  \text{ eigenvalues of } \hat{A}, \hat{B}, \hat{C}\rbrace$ and hence $Z_A\cdot Z_B = Z_C$ almost surely.} but the  condition is now only imposed on families of commuting observables that can be jointly measured. 

Families of commuting observables always have a common probability distribution (as random variables on a classical probability space). So what could possibly go wrong? One can consider an observable $\hat{A}$ once as part of a commuting family $(\hat{A}, \hat{B}, \hat{C}, \ldots)$ and once as part of a commuting family $(\hat{A}, \hat{L}, \hat{M}, \ldots)$ such that $\hat{B}, \hat{C}, \ldots$ and $\hat{L}, \hat{M}, \ldots$ are incompatible -- that is, non-commuting -- with each other. Assumption (NC) would be trivial if the observable $\hat A$ could be associated with a random variable $Z_A$, as part of the family $(Z_A, Z_B, Z_C,...)$, and another random variable $\tilde{Z}_A$ as part of the family $(\tilde{Z}_A, Z_L, Z_M, ...)$. The considered hidden-variables-schemes presuppose, however, a rigid assignment $\hat{A} \mapsto Z_A$, independent of the measurement context. In other words, $Z_A$ must be the same, whether $\hat{A}$ is measured together with $ \hat{B}, \hat{C}, \ldots$ or together with $\hat{L}, \hat{M}, \ldots$. The crucial assumption underlying the no-go theorem of \cite{Kochen:1967aa} has thus been named \emph{non-contextuality}. The upshot is that non-contextual hidden variables are incompatible with the predictions of quantum mechanics. 

A particularly nice and simple proof is due to \citet[][p. 810]{Mermin:1993aa}. It consists in the following arrangement of $3 \times 3$ observables on a 4-dimensional Hilbert space: 

$$
\begin{matrix}
\sigma_x^1 && \sigma_x^2 && \sigma_x^1\sigma_x^2 \\\\
\sigma_y^2 && \sigma_y^1 && \sigma_y^1\sigma_y^2 \\\\
\sigma_x^1\sigma_y^2  && \sigma_x^2 \sigma_y^1 && \sigma_z^1\sigma_z^2 \\
\end{matrix}
$$  
Using the standard commutation relations of the Pauli-matrices (``spin observables'') and the fact that the possible eigenvalues are ${\pm 1}$, it is easy to verify that: 
\begin{enumerate}[(a)]
\item The observables in each of the three rows and each of the three columns are mutually commuting. 
\item The product of the three observables in each of the three rows is $1$. 
\item The product of the three observables in first two columns is $1$, while the product of the right column is $-1$. 
\end{enumerate}

\noindent Thus, no consistent assignment of predetermined values to the nine observables is possible, since identity (b) would require the product of all nine values to be $+1$, while (c) would require it to be $-1$. This proves the Kochen-Specker theorem. 

\subsection{Bell}
One of the more tragic chapters in the history of quantum mechanics is that for many defenders of the supposed orthodoxy, Bell's theorem (reprinted in \citet[][ch. 2]{Bell:2004aa}) has replaced von Neumann's as the mathematical result that finally spells the dead for any ``completion'' of the quantum formalism. Certainly, the physical significance of Bell's theorem can hardly be overstated, but to understand it as just another no-hidden-variables argument is to miss the point entirely. Bell himself has addressed the misunderstanding on various occasions, for instance: 

\begin{quote} 
	My own first paper on this subject (\emph{Physics} 1, 195 (1965)) starts with a summary of the EPR argument \emph{from locality to} deterministic hidden variables. But the commentators have almost universally reported that it begins with deterministic hidden variables. (\citet[][p. 157]{Bell:2004aa})
\end{quote}
The point of Bell's theorem is not hidden variables but \emph{nonlocality} (see \cite{maudlin2014b} for an excellent discussion). Bell's analysis starts from the EPR argument that assumes locality and concludes that the quantum formalism must be incomplete. EPR did indeed attack the quantum doctrine that observables do not have predetermined values prior to measurement. In brief, they did so by noticing that, when considering two entangled systems $A$ and $B$, certain observable values of $A$ can be determined by measurements on the distant system $B$ (and \emph{vice versa}). But this would presuppose some sort of nonlocal influence \emph{unless} these values were actually predetermined, prior to the measurement on the distant system, by hidden variables (and thus only \emph{revealed} rather than  \emph{determined} by our interaction with the distant system).

Three decades later, Bell proved that even by introducing additional variables, the statistical predictions of quantum mechanics cannot be reproduced without nonlocal influences. The conclusion is thus that quantum mechanics is nonlocal, no matter what. And since a substantial amount of experimental evidence confirms the predictions of quantum mechanics, the conclusion is that any correct theory of nature is nonlocal, no matter what. Nonlocality, in other words, is not the price that we pay for introducing hidden variables. Hidden variables were Einstein's hope for avoiding the nonlocality of standard quantum mechanics, and Bell proved that this hope cannot be realized because nonlocality is a fact of nature. Hence, using nonlocality as an argument against Bohmian mechanics, or so-called ''hidden variables theories'' in general, get's the issue completely wrong. Quantum mechanics is nonlocal, and any extension of -- or alternative to -- quantum mechanics better be nonlocal as well; otherwise, it is demonstrably wrong.

\section{The message of the quantum}
So, what is the upshot of the no-hidden-variables theorems? In this section, we consider some common responses and briefly indicate why they are wrong-headed. 

\subsection{Completeness of quantum mechanics}
The no-hidden-variables theorems are usually cited in support of the claim that standard quantum mechanics is ``complete'', that is, in particular, that the wave function or quantum state  -- with its role in determining the probabilities of measurement outcomes -- represents the complete physical state of a quantum system. However, when used in this context, the traditional hidden variables program seems to commit the following mistake that Einstein warned the young Heisenberg about:

\begin{quote}I suspect that you will run into problems at exactly that part of your theory that we just talked about ... You pretend that you could leave everything as it is on the side of observations, that is, that you could just talk in the former language about what physicists observe. (Quoted after \citet[p. 89]{heisenberg2012}; translation by the authors.)\end{quote}
Indeed, the idea that physical observations must be reported in ``classical language'' (while the same language is unable to provide an objective description of the microcosm) became one of the core tenants of the so-called Copenhagen interpretation. This included the (at least tacit) assumption that the relevant observables of quantum physics are just the familiar properties known from Newtonian mechanics, or at least that the physical and ontological status of the properties, once measured, is the same as had been generally assumed in classical physics, namely that the observables refer to intrinsic properties of the physical systems. The no-hidden-variables theorems then show that the intrinsic properties of physical systems, insofar as they are captured by observables, cannot have predetermined values (unless one buys into undesirable consequences such as ``contextuality'' that seem to defeat the purpose of assuming predetermined properties.)

However, this -- orthodox -- way of reading the no-hidden-variables theorems directly runs into the two questions raised in section \ref{sec:why}: How do the quantum observables acquire definite values upon measurement? What characterizes a physical system prior to -- or better: independent of -- measurement? The ``industry of no-go theorems'' (\cite{Laudisa:2014aa}) drives us towards the negative conclusion of no predetermined values, but it does not provide an answer to these questions. Instead of this reading of no predetermined values of intrinsic properties of physical systems, there also is another, arguably more radical reading of the no-hidden-variables theorems possible: they tell us that observables do not correspond to properties of physical systems at all, so that the question of predetermined values of such properties does not even arise. This is the Bohmian reading, which then does provide an answer to these questions.      

\subsection{Metaphysical indeterminacy}
Following the lead of mainstream physics, the philosophical literature has recently developed a renewed interest in the concept of metaphysical indeterminacy, which is intended in this context to capture the idea that the values of quantum observables, prior to measurement, are not merely unknown but, in a metaphysically robust sense, unspecified. According to \cite{Calosi:2017aa}, properties of quantum systems are metaphysically indeterminate in the sense that they have a determinable property without a unique corresponding determinate. Thus, an electron, for instance, possesses a determinable property ``spin'', but its value is indeterminate until we actually measure it. 
 
In contrast, and arguing against the concept of metaphysical indeterminacy, \citet[][p. 207]{Glick:2017aa} proposes what he calls a ``sparse view'' of standard quantum mechanics: 

\begin{quote}Sparse view: when the quantum state of \emph{A} is not in an eigenstate of $\hat{O}$, it lacks \emph{both} the
determinate and determinable properties associated with $\hat{O}$.\end{quote}
Obviously, none of these views does anything to address the measurement problem, that is, to clarify \emph{how} a measurement turns an indeterminate -- or non-existent -- property of a physical system into a determinate one.  And while this is not the issue of this particular philosophical discussion, it certainly is dubious to base metaphysics on imprecise or even inconsistent physics. 

Bohmian mechanics, by contrast, shows that there is no work to do for a concept of metaphysical indeterminacy: the state of a physical system is completely and precisely determined, at any moment in time, by the actual particle positions and the wave function, fixing how the positions change in time. Furthermore, this theory supports a metaphysical view that is even sparser than the one advocated by Glick: Neither the determinate nor the determinable property associated with an observable $\hat{O}$ is part of the ontology, independent of whether or not the quantum state of a system is an eigenstate or not. The only property that particles have -- and need -- is a position in physical space (see \cite{Esfeld:2017aa} for an elaboration on a sparse ontology in that sense). 

\subsection{Quantum logic}
One of the more audacious claims in support of which the no-hidden-variables theorems are employed is that quantum mechanics compels us to give up classical logic in favor of a new quantum logic. It is easy to see where this idea comes from. If we consider the simple example of spin (for a spin-1/2-particle, to be discussed in detail in section \ref{sec:spin}) it is tempting to assign to the proposition 
\begin{quote} $q \vee \neg q$: The particle has $z$-spin up or $z$-spin down\end{quote} the truth-value \emph{true}. However, according to the doctrine $Q$, neither 
\begin{quote}
$q$: The particle has $z$-spin up
\end{quote}
nor 
\begin{quote}
	$\neg q$: The particle has $z$-spin down
\end{quote}
can be considered \emph{true} prior to a measurement or unless the particle happens to be in a $z$-spin eigenstate.

Since Quine's seminal paper ``Two dogmas of empiricism'' (\cite{Quine:1951aa}), it is widely accepted in philosophy that not even a revision of the rules of logic is out of bounds when adjusting a theoretical system to new empirical evidence, though they are the last knob to turn. In that vein, the first and most important objection to quantum logic (as a proposal for the ``true'' logic of the physical world) is not that it is \emph{a priori} absurd, but that it is hardly justified by theoretical or explanatory merits. Giving up on classical logic does nothing to address the two crucial questions formulated in section \ref{sec:why}. The various proposed systems of ``quantum logic'' are merely modeled on the standard theory and thus inherit all its problems -- including the measurement problem. In particular, changing a logical formalism does not elucidate the ontology of quantum mechanics, nor does it provide for a physical account of when and why propositions involving quantum observables acquire definitive truth values. Conversely, the example of Bohmian mechanics shows that once we have a clear ontology, and take the measurement process seriously as part of the theory, no departure from classical logic is called for. 

\section{Measurements in Bohmian mechanics: Spin}\label{sec:spin}

In this section, we explain how Bohmian mechanics treats measurement experiments, how this treatment supports doctrine Q and what the consequences for the status of observables are.

\subsection{The Bohmian treatment of the measurement process}
The solution to the measurement problem offered by Bohmian mechanics comes from a simple idea: to describe quantum mechanically also the experimental devices, since macroscopic objects are composed of microscopic objects. Thus, to describe experimental situations in Bohmian mechanics, we split the total configuration (of, in the last resort, the entire universe) into $(X,Y) \in \mathbb{R}^{3M}\times \mathbb{R}^{3(N-M)}$ where the former variable refers to the particle configuration of the investigated $M$-particle sub-system and the latter to the configuration of the environment, which includes the particles of the measurement device registering the outcomes in ``pointer positions''. Fundamentally, in the Bohmian theory, there is only one wave function, the universal wave function $\Psi=\Psi(x,y)$, guiding all the particles together (the lower case variables refer to the possible configurations -- $\Psi$ is a function on the entire configuration space --  in contrast to the \emph{actual} configurations denoted by upper case letters). However, by inserting the actual configuration of the environment at time $t$, we get a \emph{conditional wave function}, which is a function of the degrees of freedom of the sub-system only: 

\begin{equation}\label{conditionalwf}
\psi_t(x):=\Psi_t(x, Y_t).
\end{equation}

This conditional wave function is always well-defined but not very useful in practice, since it has a non-trivial dependence on the exact configuration of the environment. However, in some situations, when the universal wave function takes the form
\begin{equation}\label{effectivewf} \Psi(x,y) = \psi(x)\Phi(y) + \Psi^\perp(x,y) \end{equation}where $\Phi(y)$ and $ \Psi^\perp(x,y)$ have macroscopically disjoint support in the $y$-variables and $Y_t \in \supp \Phi(y)$, i.e. $\Psi^\perp(x, Y_t)=0 \; \forall x$, we can for all practical purposes forget about the ``empty'' wave $\Psi^\perp$ and provide an autonomous description of the subsystem in terms of the \emph{effective wave function} $\psi$, which is the Bohmian analog to the usual wave function used in textbook quantum mechanics. Now, let's consider an ideal measurement associated with an ``observable'' with eigenvalues $\alpha_1, \ldots, \alpha_n$ and corresponding eigenstates $\varphi_1, \ldots , \varphi_n$. In general, $\psi$ will be a superposition $\psi = \sum\limits_{i=1}^n c_i \varphi_i, \, c_i \in \mathbb{C}$. Under the Schrödinger evolution -- after the subsystem has coupled to the measurement device in the course of the measurement process --, the state of system + environment (ignoring again the empty part of the wave function $\Psi^\perp$) will thus have the form 
\begin{equation}
\Psi(x,y)=\sum\limits_{i=1}^n c_i \varphi_i(x) \Phi_i(y)
\end{equation}
where the environment states $\Phi_i$ are concentrated, in particular, on different pointer configurations, indicating the measurement outcomes $\alpha_i$, and have therefore pairwise disjoint supports in configuration space. Note that it is only for simplicity that we don't distinguish between the degrees of freedom of the measurement device and the rest of the universe, subsuming both in the ``environment'' ($y$-system).
However, the \emph{actual} configuration $Y$ of the universe (pointer) will lie inside only one of the branches, let's say $Y \in \supp \Phi_k$. Hence, the actual pointer configuration will indicate the measurement outcome $\alpha_k$ and the new effective (=conditional) wave function of the subsystem becomes $
\psi^{Y}(x)=c_k \varphi_k(x) \Phi_k(Y) \; \hat{=} \; \varphi_k$ after normalization. Hence, while the universal wave function always evolves according to the linear Schrödinger equation, the effective wave function automatically collapses into the eigenstate corresponding to the registered measurement result (for a detailed exposition see \citet[][ch. 9]{Durr:2009fk}).

This account has notably the following five features: 
\begin{enumerate}
   \item There never are superpositions of anything in physical space. All there is in physical space are particle configurations with always definite positions. Thus, Schr\"odinger's cat always is in a configuration of either a live cat or a dead cat. Superpositions concern only the wave function in physical space in its role to determine the trajectories on which the particles move.
  \item Consequently, quantum logic is irrelevant when it comes to an account of measurement: the particle configuration belongs unambiguously to one of the possible supports of the wave function, which in turn correspond to macroscopically different components of the experimental device, determining in this way the final outcome of the observation at hand.
    \item Nevertheless, there is entanglement in physical space: the motion of any particle depends on, strictly speaking, the positions of all the other particles in the universe via the wave function. Thus, for instance, in the double slit experiment, the motion of any particle after having passed one slit depends on the position of all the particles making up the experimental set-up, in particular on whether or not the other slit is open. This is the way in which Bohmian mechanics implements the quantum nonlocality proven by Bell's theorem. The consequence is that the trajectories of the particles often are highly non-classical.
    \item A measurement is an interaction that will in general \emph{change} the wave function of the measured system. ``Incompatible measurements'' -- corresponding to non-commuting observables -- are simply experiments in which the first measurement interaction changes the wave function in a way that influences the statistics of the second, etc.
    \item The fact that we cannot go beyond Born's rule in making predictions is explained not by any indeterminacy of the properties of the particles, or any indeterminism of the dynamics, but by the fact that we cannot have more precise knowledge of the initial particle configuration. As mentioned in section \ref{sec:why}, in Bohmian mechanics, Born's rule is \emph{derived} from the laws of motion plus a probability (more precisely: typicality) measure linked with these laws.   
\end{enumerate}
Once ``measurements'' and ``observations'' are no longer treated as primitive but as physical processes, to be analyzed on the basis of a precise microscopic theory, it turns out that the quantum orthodoxy was right about the fact that measurements do not reveal preexisting values of observables, but wrong about the idea that these observables correspond to properties of physical systems. The important contrast between classical and quantum mechanics that the no-hidden-variables theorems reveal is thus not that quantum phenomena are irreducibly random, but rather that quantum phenomena are at odds with a metaphysics of intrinsic properties that classical mechanics did not necessitate but indulge. 

\subsection{What is measured in a spin measurement?}
Let us now discuss a Stern-Gerlach spin measurement, as the simplest but maybe most instructive example of a measurement process in Bohmian mechanics. In this famous experiment, a spin-1/2-particle (originally a silver atom) is sent through an inhomogeneous magnetic field (Stern-Gerlach magnet) and then registered on a detector screen, where one observes a deviation perpendicular to the flight direction and parallel or anti-parallel to the gradient of the magnetic field. 

To describe the experiment theoretically, we consider the propagation of a concentrated wave packet

\begin{equation}\label{Phian}
\Phi_0=\varphi_0(z)\left(\alpha {1\choose
	0}+\beta{0\choose 1}\right)
\end{equation}
through an inhomogeneous magnetic field with gradient in $z$-direction. We ignore the components of the wave-function in the $x,y$-directions and the spatial spreading of the wave function, assuming that the flight time is reasonably short. A straightforward computation using the Pauli equation (which is the non-relativistic limit of the Dirac equation, describing the time evolution of a spinor-valued wave function in an external electromagnetic field) then shows that the equations for the two spin-components decouple and that each acquires a phase 
\[
\Phi^{(n)}(\tau)= \exp[\mathrm{i} (-1)^{n+1} \frac{\mu b \tau}{\hbar}z]\,\Phi_0^{(n)},
\]
where $\tau$ is the time spent in the magnetic field, corresponding to a group velocity of
\[
v_z=
(-1)^{n+1}\frac{ \mu b \tau}{m}.
\]

\noindent The inhomogeneous magnetic field thus leads to a spatial separation of the wave packets, corresponding to the spin-components: The wave packet 
$\Phi^{(1)}(t)= \alpha \varphi_1(t,z)  {1\choose
	0}$
propagates in the positive $z$-direction (in the direction of the gradient of the magnetic field) and the wave packet
$\Phi^{(2)}(t)=\beta \varphi_2(t,z) {0\choose 1}$
in the negative $z$-direction. Assuming that the two wave packets remain reasonably well localized, they will have approximately disjoint supports after a little while, that is, $\Phi^{(1)}$ is concentrated above the symmetry axis and $\Phi^{(2)}$ below. It is important to emphasize that this is purely a result of the Schrödinger (respectively Pauli) time evolution, which is part of every quantum theory, independent of interpretative issues. 

However, in Bohmian mechanics (and only there) it now makes sense to ask whether the particle moves upwards -- guided by the wave packet $\Phi^{(1)}$ -- or downwards, guided by the wave packet $\Phi^{(2)}$. In the first case, it would hit a detector screen above the symmetry axis and one says that ``the particle has $z$-spin up''; in the second case, it would hit a detector screen below the symmetry axis and one says that ``the particle has $z$-spin down''. But this is a rather unfortunate way of speaking. Spin is not a property that the particle possesses over and above its position. To ``have'' spin up or spin down means nothing more and nothing less than to be guided by the part of the wave function that corresponds to the upper or lower spinor-component (in the $z$-spin basis) -- that is, to \emph{move}, in the pertinent measurement context, in the respective way. In other words: spin is a degree of freedom of the wave function (related to its transformation under rotations) that manifests itself, under certain circumstances, in a particular kind of particle motion. As such, it belongs to the dynamical structure of the theory, not to the ontology of objects in physical space (see also \citet[][ch. 4]{Bell:2004aa} and \cite{Norsen:2014aa}).  

According to Born's rule for the particle positions, we can compute the probability of finding the particle with ``spin up'', that is, in the support of $\Phi^{(1)}$, or ``spin down'', that is, in the support of $\Phi^{(2)}$ as:

\begin{equation}\begin{split}
\IP(\text{``$z$-spin up''})=\IP(X \in \supp \Phi^{(1)}) = \int_{\supp \Phi^{(1)}} \lvert \Phi^{(1)}(t,z)\rvert^2 \dd z = \lvert \alpha \rvert^2\\
\IP(\text{``$z$-spin down''})=\IP(X \in \supp \Phi^{(2)}) = \int_{\supp \Phi^{(2)}} \lvert \Phi^{(2)}(t,z)\rvert^2 \dd z = \lvert \beta \rvert^2
\end{split}\end{equation}

\noindent Obviously, these probabilities can already be computed from the initial state, using the projections on the respective spin-components:
\begin{equation}\begin{split}
\IP(\text{``$z$-spin up''})=\langle\Phi_0 | \uparrow \rangle\langle \uparrow |\Phi_0 \rangle = \lvert \alpha \rvert^2\\
\IP(\text{``$z$-spin down''})=\langle\Phi_0 | \downarrow \rangle\langle \downarrow |\Phi_0 \rangle = \lvert \beta \rvert^2.
\end{split}\end{equation}
Finally, assigning to ``spin up'' and ``spin down'' the numerical values $\pm \frac{\hbar}{2}$, the expectation value is computed as
\begin{equation} \frac{\hbar}{2} \langle\Phi_0 |\Bigl( \lvert\uparrow \rangle\langle \uparrow \rvert -  \lvert \downarrow \rangle\langle \downarrow \rvert\Bigr)|\Phi_0 \rangle =\frac{\hbar}{2} \langle\Phi_0 |\sigma_z|\Phi_0 \rangle. \end{equation}

\noindent In standard quantum mechanics, the operator $\frac{\hbar}{2} \sigma_z$ has developed a certain life of its own as the ``spin observable''. The Bohmian analysis reveals it to be nothing more and nothing less than a convenient book-keeper of the measurement statistics (for a general discussion of observables and operators in Bohmian mechanics, see \citet[][ch. 3]{Durr:2013aa}). We should note that the example of spin is particular in Bohmian mechanics in that the statistical analysis does not require the coupling to a measurement device. It makes sense to ask whether the particle moves upwards or downwards after passing the Stern-Gerlach magnet, without considering a screen or detector in which its position is finally recorded. In many cases, though, the ``observable values'' have meaning only insofar as their are registered in some sort of ``pointer'' configuration. 

It is interesting to observe that all precise formulations of quantum mechanics, which solve the measurement problem, agree on this basic point that the measured values are \emph{produced} rather than \emph{revealed} by the interaction between system and measurement device. According to spontaneous collapse theories (such as GRW), it is the Stern-Gerlach magnet that causes the wave packets to separate and the subsequent coupling to a detector (screen) that (very very likely) causes a collapse and forces the system to go into one of the possible outcomes. According to the more sophisticated versions of Many-Worlds, it is the splitting of the wave packets in the Stern-Gerlach magnet and the subsequent interaction with a detector that leads to decoherence and a branching into ``worlds'', in which the detector has registered ``spin up'' and ``spin down'' respectively.

Only in Bohmian mechanics, however, a unique measurement outcome is determined by the initial position of the particle and the deterministic law of motion. (Collapse theories are fundamentally stochastic, while in many-worlds theories, measurements do not have unique outcomes.) That notwithstanding, it would be misleading to say that the particle possesses a predetermined spin, irrespective of the measurement context. In particular, what we end up calling the ``spin value'' is a number that encodes the result of the measurement interaction -- how the particle moves after passing the magnetic field -- by contrast to an additional physical quantity that determines it. 

\subsection{Is Bohmian mechanics ``contextual''?}
In fact, this confusion between ``predetermined outcomes'' and ``predetermined properties'' is all there is to the discussion of contextuality in Bohmian mechanics. What this theory rejects is the ``naive realism about operators'' or observables (\cite{Daumer:1996aa}) -- these unholy and categorically confused amalgams of self-adjoint operators, physical properties, and observed data points. As mentioned before, observables play no fundamental role in the theory; they merely arise, in a statistical analysis, as book-keepers of outcome statistics. Consequently, they are not properties of anything. It is simply wrong, and giving rise to further confusion, to call them ``contextual properties'' of physical systems.
In fact, different experimental setups associated with the same ``observable'' may have nothing in common besides the fact that they are associated with the same statistical book-keeping operator.

To illustrate this point, let's return to Mermin's proof of the Kochen-Specker theorem (see section \ref{sec:KS}) and focus, for instance, on the observable $\sigma_x^1\sigma_x^2$ in the upper right corner of his scheme. This observable can be trivially measured together with $\sigma_x^1$ and $\sigma_x^2$: Take two spin-1/2-particles and measure their $x$-spin separately in the way described above. Assign the value $+1$ if the particle moves in positive $x$-direction and $-1$ if the particle moves in negative $x$-direction and compute the product of the outcome values to obtain ``the value of $\sigma_x^1\sigma_x^2$''. But how to measure $\sigma_x^1\sigma_x^2$ together with $\sigma_y^1\sigma_y^2$ and $\sigma_z^1\sigma_z^2$? We have no idea, actually. In any case, one cannot simply measure the $x$-spin of particle 1 and 2 separately, as before, since this would preclude the simultaneous measurement of $\sigma_y^1\sigma_y^2$ and $\sigma_z^1\sigma_z^2$. Hence, whatever an experimentalist would have to do to perform a joint measurement of $(\sigma_x^1\sigma_x^2, \sigma_y^1\sigma_y^2,\sigma_z^1\sigma_z^2)$ -- and whatever the physical significance of this measurement might be -- it certainly requires a completely different experiment than the measurement of $(\sigma_x^1, \sigma_x^2, \sigma_x^1 \sigma_x^2)$. 

In Bohmian mechanics, the initial state (wave-function + positions) of the particles (possibly together with the initial state of the experimental setup) would determine the outcome of ``the $\sigma_x^1\sigma_x^2$-measurement'' in both experiments, but there is simply no reason why these outcomes must in every case agree. A disagreement would be troubling only if one assumed that the particles actually have a preexisting $\sigma_x^1\sigma_x^2$-property that both experiments are supposed to reveal by different methods. But this is just not the case in Bohmian mechanics. And taking the physical situation seriously, there is no reason why it should be the case in any reasonable theory. As \cite{Goldstein:2017} notes: ``If we avoid naive realism about operators, contextuality amounts to little more than the rather unremarkable observation that results of experiments should depend upon how they are performed ...''.

\subsection{Why measurements?}
Nonetheless, since, according to Bohmian mechanics, the outcome of any measurement is determined by the initial state of the system (or at least of system  + apparatus), the measurement outcome does reveal a certain amount of information about the state of the system prior to measurement. In fact, in some cases, the Bohmian theory allows us to infer significantly more information about the measured system than standard quantum mechanics does. If we consider, for instance, a $z$-spin measurement on a particle in the spin state $\frac{1}{\sqrt{2}}(|\uparrow_z\rangle + |\downarrow_z\rangle)$ and assume that the setup is reasonably symmetric about the incident axis, we can infer from the ``no-crossing property'' of Bohmian trajectories that if a particle hits the screen above/below the symmetry axis (corresponding to $z$-spin up or $z$-spin down, respectively), it's initial position must have been above/below the symmetry axis as well. 

In general, though, a quantum experiment provides more information about the state of the system \emph{after} the measurement process. In particular, if we perform an ideal (projective) measurement and find a non-degenerate eigenvalue $\alpha$ of some observable $\hat A$, we know that the effective quantum state of the system after the measurement is the corresponding eigenstate $\psi_\alpha$. According to Bohmian mechanics, this quantum state is an objective physical degree of freedom of the system (in accordance with the PBR-theorem \citep{pusey.etal2012}), providing statistical information about the particle configuration and determining its state of motion. It is thus highly informative about the future behavior of the system. Note, however, that it would be wrongheaded to interpret the effective quantum state as an additional intrinsic property of the particles, a) because one can, in general, assign a wave function only to the subsystem as a whole but not to each particle individually (non-separability) and b) because the effective wave function depends -- implicitly -- on the universal wave function and the configuration of all the other particles in the universe (cf. equations \eqref{conditionalwf} and \eqref{effectivewf}).

Orthodox quantum mechanics agrees that a measurement provides, in general, more information about the post-measurement state of the system, but would, strictly speaking, disagree on what the information is actually about. The disagreement can be summarized as follows: According to Bohmian mechanics, the ``observable values'' are best understood as encoding information about the quantum state (i.e. the dynamical state) of the system, while according to standard quantum mechanics (or at least most versions thereof), the quantum state is understood as encoding information about the observable values. What makes the Bohmian view more coherent is the fact that the observable values per se  -- in contrast to the quantum state -- \emph{have no causal role within the theory}.\footnote{Except maybe for conserved quantities, but even those get their physical significance mostly in the ``classical limit''.}  To appreciate this point, it might be helpful to engage in a little thought exercise: Suppose we write down some abstract self-adjoint operator $\hat A$ on a Hilbert space and tell you that a certain physical system (an electron, let's say) has the value $\alpha$ of this ``observable''. What information have we actually given you about the world? How would you (or any other physical system) have to interact with the electron to ``notice'' that it has the $\hat A$-value $\alpha $ rather than $\alpha'$? Try to answer these questions by taking the physical theory seriously, whatever you consider quantum theory to be. 

 \section{Are observables observable?}\label{sec:positions}
The suggestive but misleading terminology of ``measuring an observable'' has not only lead to a naive realism, but also to a naive empiricism about observables in quantum mechanics. It is usually taken for granted that all empirical data underlying quantum physics consist in measured values of observables, represented by -- or corresponding to -- self-adjoint operators. Against this backdrop, our previous analysis seems to lead to a certain dilemma. Since the measured values of quantum observables are emergent in a measurement process, they must emerge from an underlying ontology that is not itself characterized in terms of definite values of quantum observables. This seems to leave us with two possible options:
 
\begin{enumerate} 
\item The physical properties are not observable. 

	\item The physical properties are a small subset of the observables (small enough to avoid the no-hidden-variables results).
\end{enumerate} 
Both options invite criticism. In the first case, the underlying ontology would have no direct empirical basis. The second option is open to the charge of arbitrariness, as it seems to reify some observable properties but not others that have the same empirical status. In fact, both lines of attack are occasionally used against Bohmian mechanics, the first in form of the claim that ``the Bohmian trajectories cannot be observed'', the second in form of the question  ``why take the position as your `hidden variable' and not something else?''. While some interesting remarks could be made in response to these objections, we want to take a step back and question the basic assumption that the ``observables'' are somehow \emph{a priori} given as fundamental objects of empirical observation.   

Consider the following image (Fig. 1) from an original Stern-Gerlach experiment, reported as the first experimental observation of a ``quantized direction (Richtungsquantelung)'' of the angular momentum / magnetic moment of atoms in an external magnetic field \citep{gerlach.stern1922}. Should we say that what was actually observed in this experiment -- what the empirical data consists in -- is the particles' spin?  

\begin{figure}[ht]
	\begin{center}
	\includegraphics[width=300px]{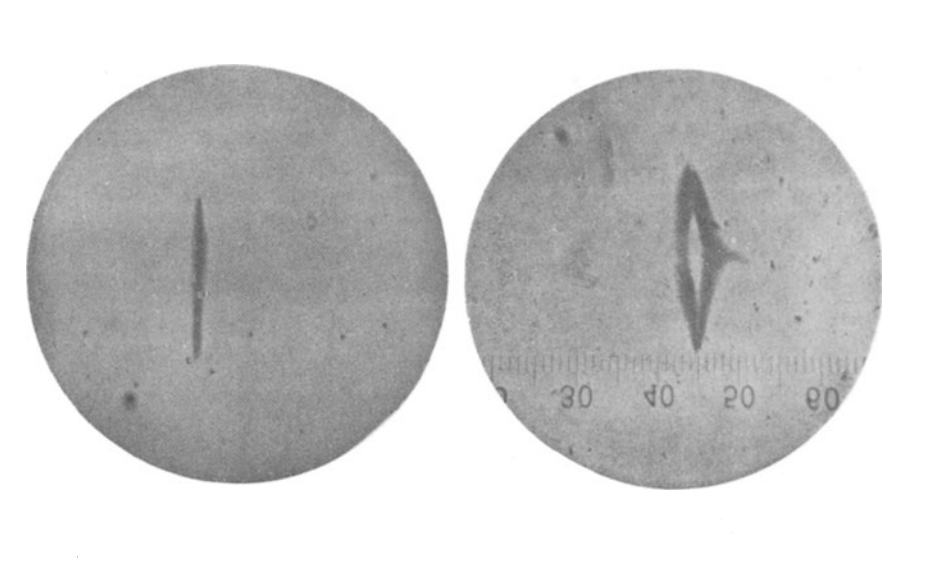}
	\caption{Pattern created by a ray of silver atomes in the original Stern-Gerlach experiment:  left without, right with magnetic field.}
	\end{center}
	\end{figure}

\noindent Evidently, our more immediate observation is that of dark marks on a screen, the ``non-classical two-valuedness'' being manifested in the distinct separation of the arcs on both sides of the symmetry axis, when the magnetic field is turned on. And evidently, the statistics of ``spin up''  and ``spin down'' (deviation to the right / left) alone are too coarse-grained to capture all observable details of the pattern.  

But this now puts the orthodox view in a predicament. Either quantum mechanics could describe the experiment, in more detail, as series of position measurements (the points of impact of the atoms building up the pattern on the screen); then the spin observable is redundant or, at least, derivative upon the observable ``position''. Or standard quantum mechanics somehow compels us to describe this experiment as a measurement of ``spin''. Then the theory is empirically incomplete, since it cannot -- even statistically and in principle -- account for all observable details of the experimental outcome.

In general, all that we observe are the positions of discrete objects and the change of these positions. Of course, there is more to these discrete objects than their mere positions, that is, spatial relationships and change of these relationships. They notably have different colors, which makes it possible to discern them in perception. But color perception is not an observable that figures in any physical theory, and the quantum observables do not help us to come up with an account of color perception. In electromagnetism, ``colors'' are identified with certain wavelengths in the electromagnetic field. But the electromagnetic field should be first and foremost understood in terms of its role for the motion of particles (and be it particles in our visual receptors). In other words, we do not observe fields, but only certain patterns of motion that we explain and calculate in terms of fields (cf. \citet{lazarovici2018}). For classical electrodynamics, even a field free formulation is available, namely the one of \cite{Wheeler-Feynman:1945aa}, which may have a number of drawbacks, but certainly does not fail for the reason that it denies alleged field observations. By the same token, even in the case of the gravitational waves detected by LIGO in 2016, all the evidence is evidence of change in the relative positions of particles, which is then mathematically described in terms of a wave rippling through the gravitational field. 

\citet[][p. 166]{Bell:2004aa} considered it to be the first and foremost lesson of Bohmian mechanics that \begin{quote} in physics the only observations we must consider are position observations, if only the positions of instrument pointers. It is a great merit of the de Broglie--Bohm picture to force us to consider this fact. If you make axioms, rather than definitions and theorems, about the `measurement' of anything else, then you commit redundancy and risk inconsistency.\end{quote}
This crucial point applies to the whole of physics. Also in classical mechanics, we do not observe mass when we observe gravitational attraction, and we do not literally see angular momentum when we notice the regular motion of the moon around the earth. What we observe is just that: certain regularities in the motion of matter, which are captured by the dynamical structure of the theory.

Hence, even in classical physics, quantities like energy, momentum, angular momentum, etc. get their meaning and relevance from what they tell us about the way matter moves. The same applies also to the classical parameters of mass and charge. Ernst \citet[][p. 241]{Mach:1919aa} highlighted this issue when he emphasized in his comment on Newton's \emph{Principia} that ``The true definition of mass can be deduced only from the dynamical relations of bodies''. In Bohmian mechanics, then, the way matter moves is encoded in the wave function, making all additional properties unnecessary or redundant. (Mass and charge, as well, are best understood as situated on the level of the wave function, instead of being intrinsic properties of the particles, see most recently \cite{Pylkkanen:2014aa} and \cite{Esfeld2014}.) This is the basic reason why Bohmian mechanics endorses doctrine Q. In that respect, the lesson of the no-hidden-variables theorems is that in quantum mechanics, one cannot treat the observables as properties of the physical systems, whereas in classical mechanics, one does not run into a problem with the physics if one regards quantities like energy, momentum, angular momentum, etc. as properties of the physical systems (although there is no cogent reason to do so in classical physics either). 

Any quantum theory that admits what is known as a primitive ontology of matter in physical space privileges position -- be it the position of permanent particles as in the Bohm theory, be it the value of the density of matter at the points of physical space as in the GRWm theory, be it single events (flashes) occurring at some points of space as in the GRWf theory (see \cite{Allori:2008aa}). In all these theories, the quantum observables are construed on the basis of the positions of objects, namely in terms of how these positions behave in certain experimental contexts. Also in the many worlds theory, which does not recognize a primitive ontology of matter in physical space, but proposes an ontology in terms of the universal wave function, position is privileged: it is the position basis in which the wave function decoheres, splitting into different branches, which constitute ``many worlds'' on this view.

 \section{Conclusion}\label{sec:concl}

What we perceive with the naked eye are the positions of macroscopic objects. But we know from scientific experience that the macroscopic objects are composed of discrete microscopic objects. If the macroscopic objects have precise positions when we observe them, so do the microscopic objects. There is no coherent theory of a magic power of the mind to change macroscopic objects in such a way that they acquire positions only when a being with a mind perceives them. So the macroscopic objects better have positions independently of someone observing them. If not the moon, so surely the desk in my office is there also when I do not observe it. But then it follows that also the microscopic objects that compose these macroscopic objects do have positions independently of them being observed. Again, there is no coherent theory according to which there is something special about the microscopic objects that compose my desk and the like. So the conclusion is that the microscopic objects \textit{tout court} have a position independently of them being observed.      

Bohmian mechanics shows how to build a quantum theory on this simple and obvious reasoning. Superpositions then concern only the parameter that encodes the dynamics of the particles, namely the wave function, but not the particles themselves. This insight is the key to answering the two questions raised at the beginning of this paper and to avoid all the puzzles of standard quantum mechanics, such as notably the measurement problem. However, as it is trivial that physical objects have positions, so it is trivial that in order to access these positions, we have to interact with these objects and thereby change their positions. Generally speaking, for one particle configuration, say a macroscopic object, to contain information about the positions of other particles, there must be a correlation between them, which is, furthermore, reliable in the sense of being reproducible. This applies in particular to correlations between particle configurations in human brains and particles outside the brains, assuming that all the perceptual knowledge that persons acquire passes through their brains.

Hence, for reasons stemming from the very way in which we acquire knowledge about the natural world, a limited accessibility of physical objects is to be expected. In that sense, classical mechanics is an idealization, and quantum mechanics brings out that limitation on our knowledge. In Bohmian mechanics, this is done in the theorem of ``absolute uncertainty'' (\citet[][ch. 2]{Durr:2013aa}), stating that we cannot have more information about the actual particle configuration of a sub-system than what is provided by the $|\psi|^2$-distribution in terms of its effective wave function. That notwithstanding, there is, of course, no question of an \textit{a priori} deduction of this theorem -- or the Heisenberg uncertainty relations -- from general conditions of our knowledge. It is just that some principled limit on our knowledge of particular matters of fact -- such as initial conditions of physical systems -- is to be expected.

If the evolution of the physical systems is highly sensitive to slight variations in their initial conditions, as is the case with quantum systems, it then follows that in general we can only make statistical predictions about the behavior of ensembles of physical systems prepared under the same conditions, but not predictions about the evolution of an individual system, although the laws of motion that govern the evolution of these systems can be fully deterministic (cf. \cite{Oldofredi:2016aa}). Again, classical mechanics is deceptively generous in this respect, and quantum mechanics brings out a fact that turns out to be trivial upon reflection (and actually comes out already in classical statistical mechanics): deterministic laws require a probability measure to yield predictions, which then are statistical. However, all these are facts about epistemology, the theory of knowledge -- as the word ``uncertainty relations'' clearly brings out --, and not about ontology, that is, about what there is in the world.      
 
Nonetheless, the no-hidden-variables theorems have a great merit: they tell us that a sparse ontology of positions is not just good metaphysics, but strongly suggested by our best theory of physics. In classical mechanics, one can attribute dynamical parameters and all sorts of  ``observables'', which are functions of the particle positions and momenta, as intrinsic properties to the particles. This does not lead to conflict with the phenomena because the active role of the measurement process -- both in producing the data and in changing the state of the measured system -- can be usually neglected in the classical regime. In quantum mechanics, as we have seen, the situation is markedly different. The moral then is of course not that there is nothing if one cannot go from observables to ontology, but that one has to start with conceiving a -- provisional, hypothetical -- ontology for whose evolution the dynamical parameters then are formulated. The guideline for this is the experimental evidence together with the coherence and explanatory fruitfulness of the proposed ontology. Bohmian mechanics shows how the simplest suggestion in that respect -- the evidence of discrete objects and their composition by discrete microobjects suggesting to try out a particle ontology -- can go through also in the quantum case and yield all the explanations that one can reasonably demand.

In a nutshell, the lesson of the no-hidden-variables theorems is that it is \emph{position only} when it comes to the ontology of the physical world, and Bohmian mechanics teaches us how to do physics on that basis (see \cite{Esfeld:2017aa} for a general treatment of that insight from classical mechanics to QFT). Note that this is not about classical vs. quantum. The ontology neither is classical nor quantum. The dynamics may be classical (as in local field theories) or quantum. What a quantum dynamics has to be subsequent to Bell's theorem is clearly brought out by the nonlocality implemented in the Bohm theory. There is no \emph{a priori} explanation of why the dynamics of the world is nonlocal. But this nonlocality fundamentally deviates from the ideas that drive classical field theory, showing a profound interconnectedness (holism) of the things in the universe.    

\bibliographystyle{apalike}
\bibliography{references_unobservables}

\end{document}